\documentclass[12pt,preprint]{aastex}
\usepackage{url}
\usepackage{natbib}
\usepackage{aas_macros}
\usepackage{hyperref}
\usepackage[usenames, dvipsnames]{color}
\shorttitle{Origin of Deep Acoustic Sources Associated with Solar Magnetic Structures}

\title{Origin of Deep Acoustic Sources Associated with Solar Magnetic Structures}
\author{I.~N. Kitiashvili$^{1}$, A.~G. Kosovichev$^{2}$, N.~N. Mansour$^{1}$, A.~A. Wray$^{1}$, T.~A. Sandstrom$^{1}$}
\affil{$^1$NASA Ames Research Center, Moffett Field, Mountain View, CA 94035, USA}
\altaffiltext{1}{e-mail: irina.n.kitiashvili@nasa.gov}
\affil{$^2$New Jersey Institute of Technology, Newark, NJ 07102, USA}

\begin{document}
\begin{abstract}
It is generally accepted that solar acoustic ($p$) modes are excited by near-surface turbulent motions, in particular, by downdrafts and interacting vortices in intergranular lanes. Recent analysis of Solar Dynamics Observatory data by \cite{Zhao2015} revealed fast-moving waves around sunspots, which are consistent with magnetoacoustic waves excited approximately 5~Mm beneath the sunspot. We analyzed 3D radiative MHD simulations of solar magnetoconvection with a self-organized pore-like magnetic structure, and identified more than 600 individual acoustic events both inside and outside of this structure. By performing a case-by-case study, we found that surrounding the magnetic structure, acoustic sources are associated with downdrafts. Their depth correlates with downdraft speed and magnetic fields. The sources often can be transported into deeper layers by downdrafts. The wave front shape, in the case of a strong or inclined downdraft, can be stretched along the downdraft. Inside the magnetic structure, excitation of acoustic waves is driven by converging flows. Frequently, strong converging plasma streams hit the structure boundaries, causing compressions in its interior that excite acoustic waves. Analysis of the depth distribution of acoustic events shows the strongest concentration at 0.2 -- 1~Mm beneath the surface for the outside sources and 2.5 -- 3~Mm for the excitation event inside the structure. 
\end{abstract}

\section{Introduction}

The turbulent nature of solar convection leads to excitation of numerous acoustic waves observed in the photosphere with a characteristic frequency of 3~mHz (known as 5-minute oscillations) in the quiet-Sun regions and 5~mHz (or 3-minute oscillations) in sunspot umbrae, with transition to 5-min oscillations in sunspot penumbrae \citep{Beckers1972}. The observed signals represent superposition from individual sources in the turbulent magnetoconvection. The origin and characteristics of these oscillations have been investigated mostly in terms of resonant modes, both observationally \citep[e.g.][]{Renaud1999,Schmidt1999,Komm2000} and theoretically \citep[e.g.][]{Stein1967,Ulrich1970,Goldreich1990,Goldreich1994,Musielak1994,Georgobiani2000}. Observations of individual acoustic events \citep{Goode1992,Strous2000,Hoekzema2002,BelloGonzalez2010a} in the intergranular lanes suggest that the excitation mechanism can be associated with strong local cooling and downflows \citep{Rast1995,Rast1999}. Different authors have estimated the acoustic source depth. For instance, it was suggested that the sources can be located at a depth of $75\pm50$~km by comparison of the power spectra of velocity and intensity oscillation modes from SoHO/MDI \citep{Nigam1999a}; 200~km or less from comparison of 1D piston-like simulations with phase shifts observed in an absorption line profile \citep{Goode1992}; 300~km or deeper by \cite{Kumar1991} from analysis of observed and modeled oscillation power spectra; 120 -- 350 km and 700 -- 1050 km for dipole and quadrupole acoustic sources by modeling the asymmetry of modal lines in the power spectrum \citep{Kumar1999a}. 

Our previous 3D radiative hydrodynamic simulations of quiet-Sun regions demonstrated that acoustic events could be associated with ubiquitous interaction of vortices formed primarily in the intergranular lanes \citep{Kitiashvili2011}. Such vortices are found in high-resolution observations, \citep[e.g.][]{Bonet2008,Bonet2010}. In these hydrodynamic simulations the acoustic sources also were found in the subphotospheric layers at a depth of 100 -- 300~km. Another different mechanism of wave excitation by small granule collapse was suggested by \cite{Skartlien2000} who estimated the source location in the height range above a cooling layer 500~km thick.

The detection of the fast-traveling waves in the frequency range of 2.5 -- 4.0~mHz around sunspots with estimated source location approximately 5 Mm beneath a sunspot \citep{Zhao2015} recently was supported by the simulations of \cite{Felipe2017} who performed a parametric study by placing artificial localized sources at various depths and compared the surface wave speed with the observations. However, due to uncertainty of the helioseismology analysis, the location of the acoustic sources relative to the subsurface magnetic structure, as well as the mechanism of wave excitation in the deep layers, remained unclear.

These results raise questions about possible mechanisms of wave excitation in subsurface layers in the presence of magnetic fields and about the distribution of acoustic sources in the presence of magnetic structures. A natural way to investigate this problem is to analyze 3D realistic MHD simulations, which model the turbulent plasma dynamics, radiation, and magnetic effects with a high degree of realism, allowing one to look at the physical processes that are hidden from direct observations. In this paper, we analyze the simulation data of \cite{Kitiashvili2010}, which reproduced spontaneous formation of a dynamically stable self-organized magnetic structure. The structure resembles the concentrations of magnetic fields observed as solar pores. They are more dynamic and smaller than sunspots but nevertheless allow insight into the nature of deep acoustic sources. In this paper, we identify and examine acoustic waves in the simulation domain and consider possible sources of their excitation as well as effects of magnetic fields.

The paper starts with a brief overview of the MHD simulation setup (\S\ref{sec:setup}) and a description of the formation and dynamics of the pore-like magnetic structures. Then we describe the properties of the acoustic waves identified both outside and inside a pore (\S\ref{sec:pore}). A statistical analysis of the acoustic sources is presented  \S\ref{sec:pore-stat} , followed by discussion of the analysis results and conclusion in \S\ref{sec:conclusion}.

\section{Numerical simulations}\label{sec:setup}

We use 3D radiative MHD simulations of the upper layers of the solar convection zone and the low atmosphere obtained with the StellarBox code \citep{Wray2015,Wray2018}. The StellarBox code is a high-order fully compressible MHD solver. It uses a Large Eddy Simulation (LES) formulation for subgrid-scale turbulence modeling and includes a fully coupled radiation solver, assuming Local Thermodynamic Equilibrium (LTE). The radiative transfer between fluid elements is calculated for four spectral bins using the long-characteristics method. The boundary conditions are periodic in horizontal directions. The bottom and upper boundaries are closed for flows ($\rm{U_z=0}$) and open for the radiative energy flux. 

\begin{figure}[h]
\begin{center}
\includegraphics[scale=1]{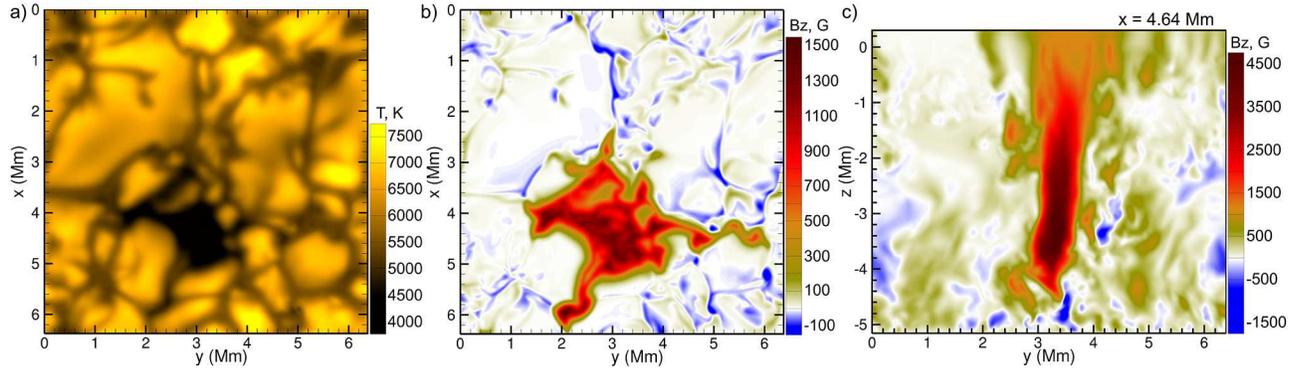}
\end{center}
\caption{Temperature (panel a) and vertical magnetic field (panel b) distributions in the photosphere in vicinity of the pore-like magnetic structure. Panel c) shows a vertical slice of the vertical magnetic field through the structure at $x = 4.64$~Mm).
\label{fig:TBz}}
\end{figure}

In the simulations, we first calculated a fully developed regime of hydrodynammic solar convection using a standard solar model for the initial conditions. Then, a uniform vertical magnetic field, $Bz^0=100$~G, was added to the simulation domain. The boundary conditions for the magnetic field maintained only the mean magnetic field strength, with no artificial structure imposed. The simulations revealed spontaneous formation of a compact, 1 -- 2~Mm in size pore-like structure with a magnetic field strength of $1.5-1.7$~kG at the photosphere and about 6~kG in deeper layers (Fig.~\ref{fig:TBz}). The self-organization process responsible for the development of the pore-like structures is described by \cite{Kitiashvili2010}. In the simulations with a domain size of $6.4\times 6.4$~Mm, one structure was formed. The simulation domain includes a 5~Mm thick upper part of the convection zone and about 300~km above the photosphere. The grid-spacing was $\Delta x = \Delta y = \Delta z = 25$~km. To model unresolved turbulent dissipation we use the Smagorinsky sub-grid turbulence model. Our analysis showed that the compact magnetic structures are maintained by strong converging downdrafts, resembling Parker's cluster sunspot model \citep{Parker1979f} and local helioseismology inferences \citep{Zhao2010}, which reveal filamentary subsurface structuring of a sunspot and converging flows. However, we note that the simulated magnetic structures are much smaller and more dynamic than sunspots. The properties of these self-formed magnetic structures correspond to pores. Another limitation of the simulations is that waves are reflected from the bottom boundary. Therefore, we primarily focus on the wave dynamics in the 3~Mm deep subsurface layer and exclude from analysis wave excitation events for which the excited wave interferes with waves reflected from the bottom boundary.

\section{Acoustic waves in subsurface layers}\label{sec:pore}

The simulated magnetic structures are evolving and very dynamic. The complexity of the dynamics and the small size of the computational domain prevent us from performing a cross-correlation time-distance analysis similar to the observational analysis of \cite{Zhao2015}. Instead, the simulation data allow us to study individual wave fronts and the corresponding excitation events in the subsurface layers. To capture the propagating acoustic wave fronts, we calculate the time-difference of the gas pressure fluctuations, normalized by the square root of the mean pressure suggested by \cite{Stein1967}: $p^*=\Delta_t(p')=\Delta_t(\frac{(p-\bar p)}{\sqrt{\bar p}})$, where $\Delta_t$ denotes the time-difference between the consequent moments of time.  These data allow us to visually detect propagating acoustic waves and to identify approximate locations of their sources by tracking the wave front locations back in time. This way, we estimated locations for more than 600 acoustic events, both inside the pore and in the surrounding region.

The analysis is performed event-by-event. The typical uncertainty of determination of the source location is about 100~km. But sometimes the uncertainty can be very high (up to 0.5 --1~Mm), even in cases of clear wave front identification, for example when several interacting downdrafts that are close to each other drive excitation of several waves, making it impossible to identify individual sources. Often these excitation events produce joint wave fronts. Therefore, all excitation events with uncertainty of source locations of more than 100~km are removed from the statistical analysis. Also, we excluded from the statistical analysis most of sources located below a depth of 3~Mm because of interference with waves reflected from the bottom boundary.

\subsection{Acoustic waves in vicinity of a pore}

\begin{figure}
\begin{center}
\includegraphics[scale=0.95]{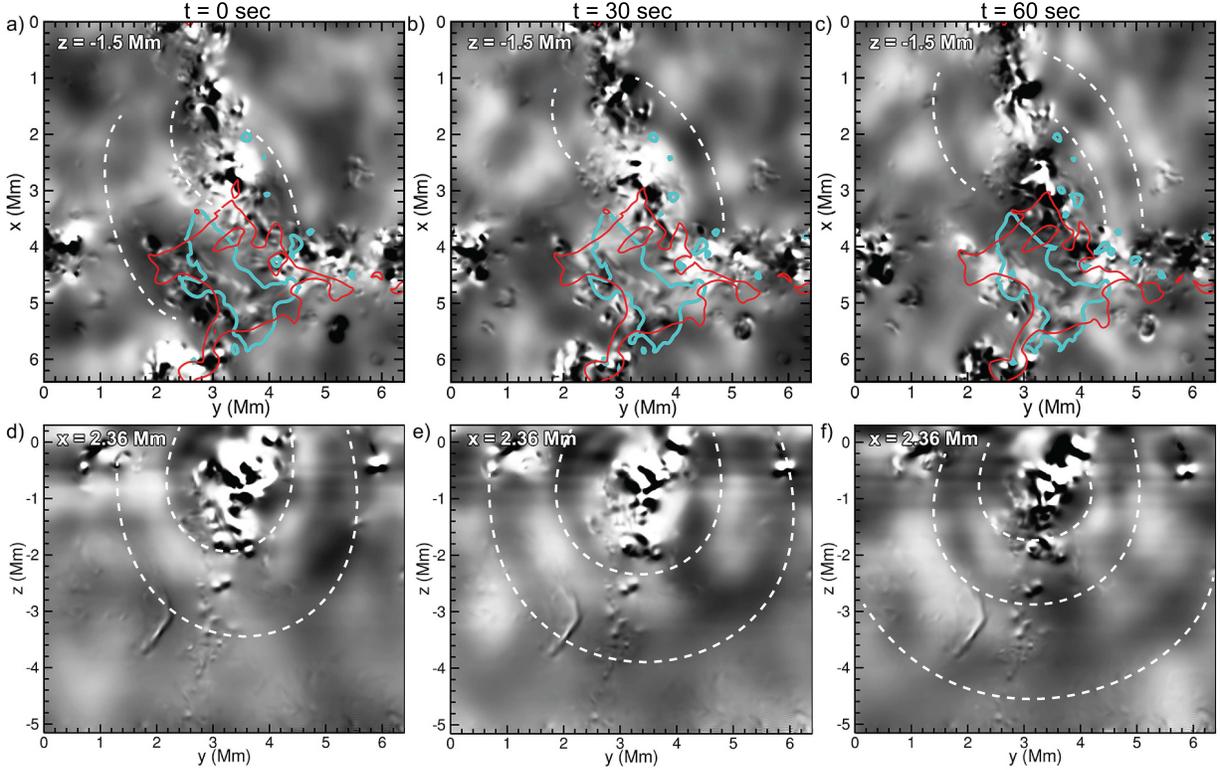}
\end{center}
\caption{Example of propagation of an acoustic wave excited in the pore vicinity showing the time-difference of gas pressure fluctuations, $p^*$, in a horizontal plane (at depth $z=-1.5$~Mm) and vertical slices ($x=2.36$~Mm). Gray colorscale is saturated at $\pm 30$~g$^{1/2}$cm$^{-1/2}$s$^{-2}$. Color contours in panels a-c) correspond to the vertical magnetic field strength, $B_z=500$~G in the photosphere (red) and at the depth of 1.5~Mm (blue). The time difference between the vertical columns is 30~sec. The time for panels a) and d) correspond to Fig.~\ref{fig:TBz}. White dashed curves indicate the locations of the wave fronts.
\label{fig:wave}}
\end{figure}

\begin{figure}
\begin{center}
\includegraphics[scale=0.41]{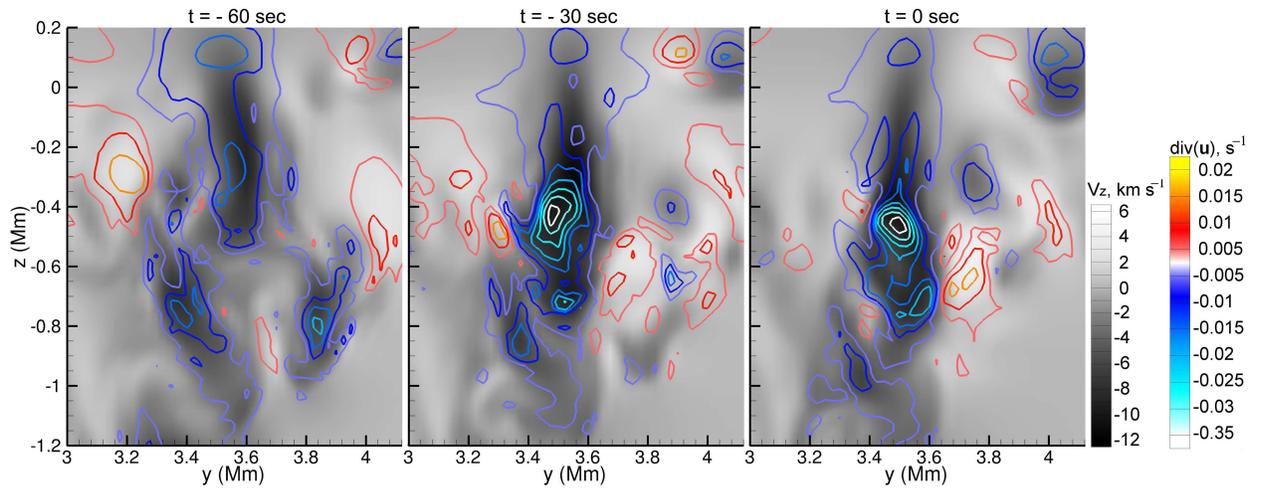}
\end{center}
\caption{Dynamics of the turbulent flow preceding the acoustic-wave excitation event shown in Figure~\ref{fig:wave}. Panels show a series of vertical slices of vertical velocity and velocity divergence with 30-sec cadence, which illustrate local variations associated with the downdraft dynamics that lead to excitation of the acoustic wave. The last moment of time (rightmost image) corresponds to Figure~\ref{fig:wave} a and d. Background gray-scale images show the vertical velocity distribution. Contour lines correspond to the velocity divergence. 
\label{fig:zoom}} 
\end{figure}

As discussed above, acoustic waves are excited both outside and inside the magnetic structure. Acoustic events outside the pore are ubiquitous and usually associated with downflows. Figure~\ref{fig:wave} illustrates an example of propagating acoustic waves excited outside the pore structure. The wave fronts are shown in a horizontal plane at depth $z=-1.5$~Mm below the photosphere (upper row) and in a vertical slice at $x=2.36$~Mm, with a cadence of 30~sec between the snapshots. For acoustic wave visualization, we use time-difference of the normalized gas pressure perturbation, $p'$. The isocontours in the top panels of Figure~\ref{fig:wave}{\it a-c}  correspond to 500~G vertical magnetic field at the photosphere, $z=0$ (red contours), and at depth --1.5~Mm (blue contours). Note that the chosen horizontal and vertical slices do not cross the acoustic source location, because otherwise strong turbulent flows would make it difficult to identify and track the wavefront. The source in this case is located near the pore boundary, which occurs quite frequently. Figure~\ref{fig:zoom} shows in detail the preceding dynamics in the vicinity of the downdraft associated with the acoustic event shown in Figure~\ref{fig:wave}. It reveals a strong local increase in the negative plasma divergence in the downdraft before the excitation of acoustic waves. Also, we found that the identified acoustic source at a depth of about 600~km below the photosphere is transported by the downflows into deeper layers with a speed of about 1.5~km/s.

\begin{figure}
\begin{center}
\includegraphics[scale=0.94]{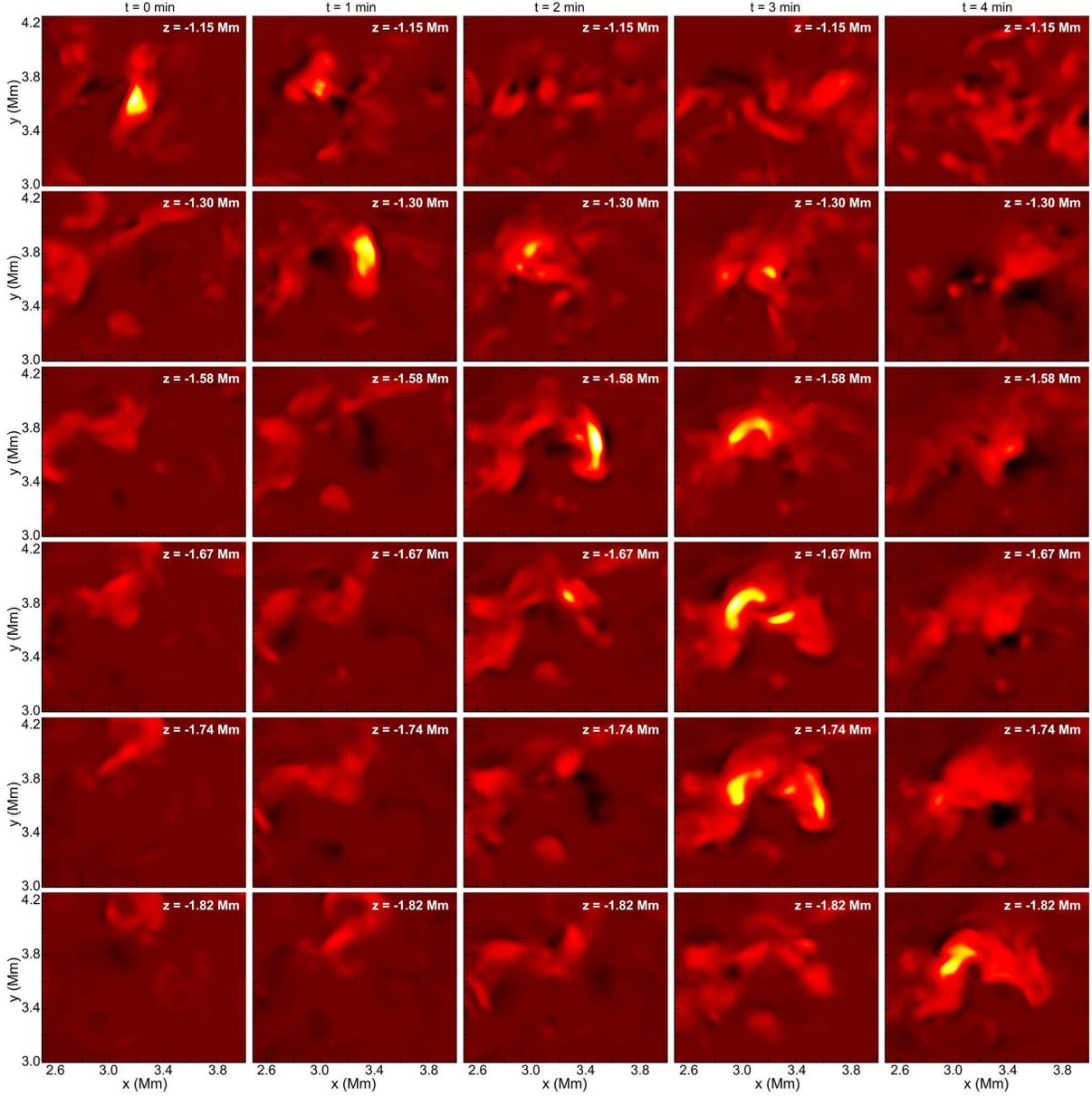}
\end{center}
\caption{Evolution of the acoustic wave energy flux density with depth (columns) and time (rows). The time-difference between snapshots is 1~min. The color-scale is in the range from $-10^{11}$ to $3.7 \times 10^{11}$~erg cm$^{-2}$~s$^{-1}$.
\label{fig:2D_F}}
\end{figure}

In general, the acoustic sources are often associated with braking of the downdrafts due to their interaction with the denser plasma of the deep layers. The shape of wavefronts is often distorted by surrounding turbulence and sometimes can be stretched and inclined along the downdrafts. In the case of multiple acoustic sources located near each other (when several interacting downdrafts merge together), usually we can identify a joint wavefront that behaves as if from a spatially distributed single source. Because of the complicated local downdraft dynamics, to study propagation of source-related disturbances we consider the time-space evolution of the acoustic wave energy flux density induced by turbulent flows, defined as a $F_{ac}=p'V_z$ \citep{Eckart1960,Musielak1994}. Figure~\ref{fig:2D_F} shows the downward propagation of turbulent disturbances associated with the acoustic flux from a source depth, $-1.15$~Mm to $-1.82$~Mm with a speed of 3 -- 4~km/s. These disturbances accompany the downdrafts and reveal swirling motions. The yellow-white color corresponds to flux values above $2 \times 10^{11}~{\rm ergs~cm^{-2} s^{-1}}$. The mean acoustic wave energy flux at the solar photosphere is about $5 \times 10^8~{\rm ergs~cm^{-2} s^{-1}}$ . This is one order of magnitude higher than the flux estimated from a mixing-length solar model by \cite{Musielak1994}. 

\subsection{Acoustic waves inside the magnetic structure}

\begin{figure}
\begin{center}
\includegraphics[scale=0.5]{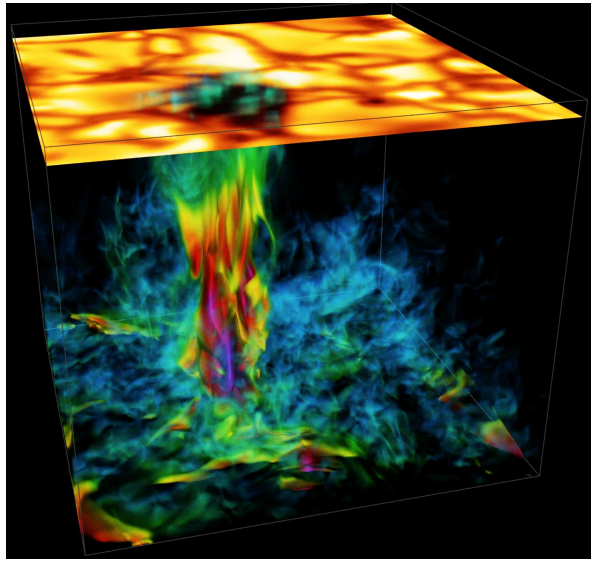}
\end{center}
\caption{3D rendering of the magnetic field strength shows the filamentary subsurface structure of the pore. Interaction of the magnetic bundles with surrounding converging flows is a possible mechanism of acoustic wave excitation inside the magnetic structure. The field strength beneath the surface reaches 6 -- 7~kG  (magenta color). The color scale is partially transparent for weaker fields. Horizontal plane shows the temperature distribution in the photosphere. 
\label{fig:3D}}
\end{figure}

\begin{figure}
\begin{center}
\includegraphics[scale=1.3]{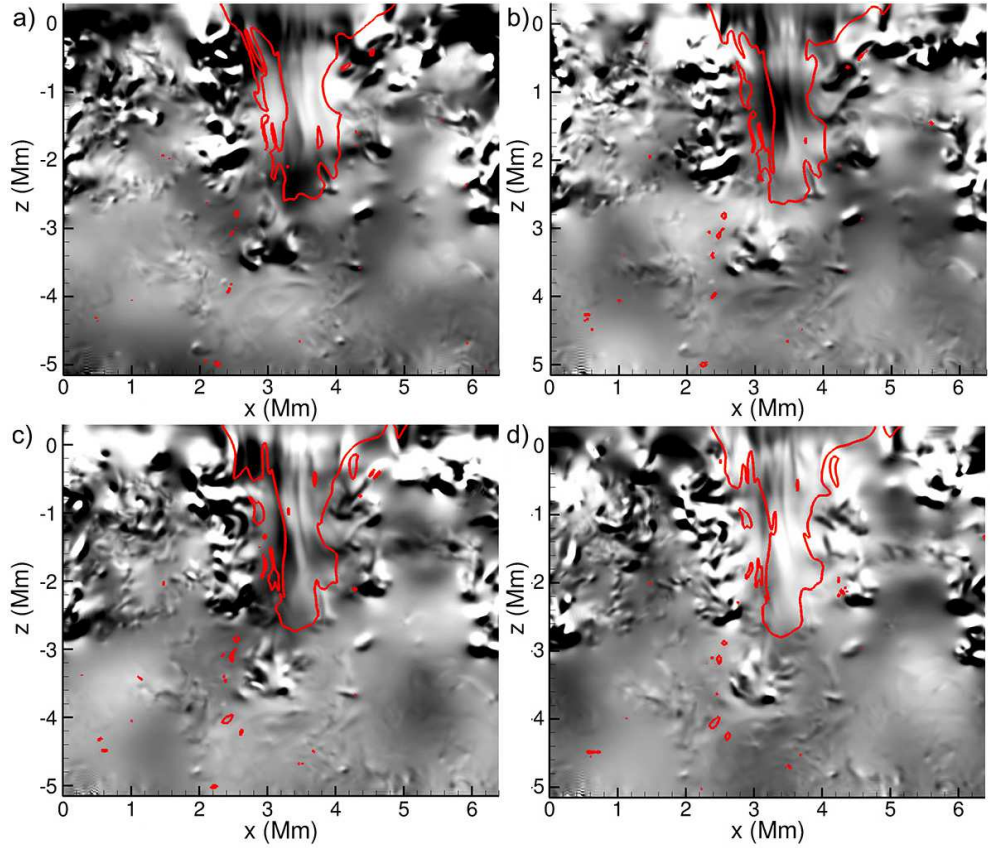}
\end{center}
\caption{Acoustic wave excitation in the pore shown as a sequence of time-differences of the normalized gas pressure fluctuations, $p^*$, through the pore (gray scale). Red isocontours of plasma $\beta = 1$ indicate the pore "boundary". The time-difference between the snapshots is 1~min.
\label{fig:2D_wave}}
\end{figure}

\begin{figure}
\begin{center}
\includegraphics[scale=1.5]{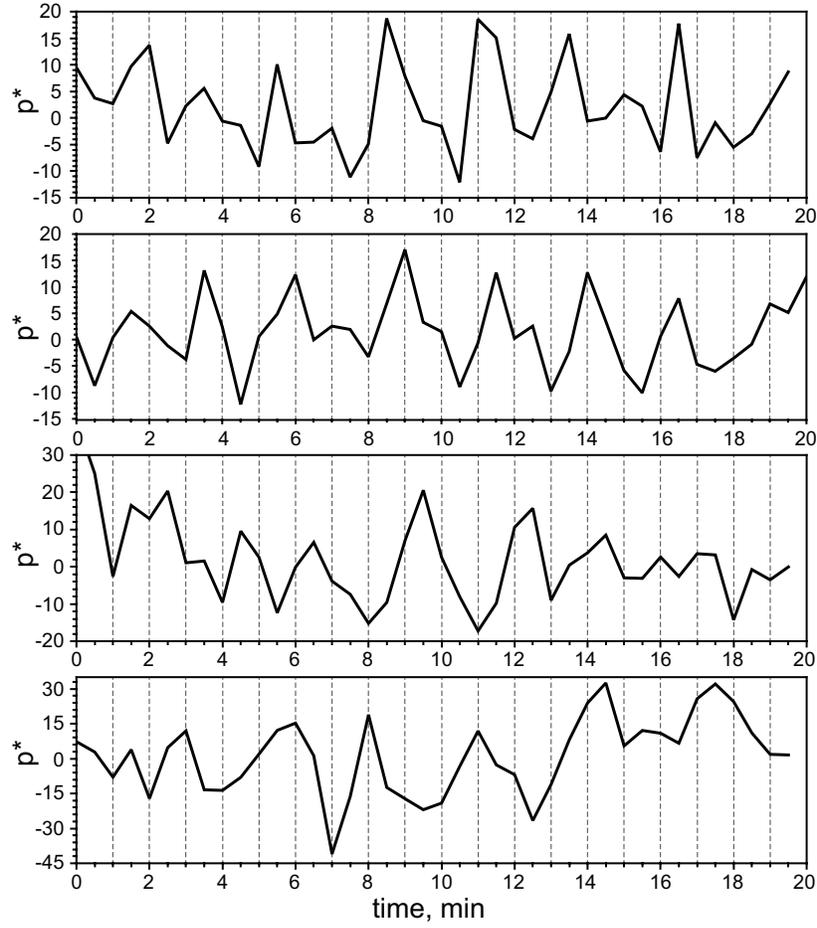}
\end{center}
\caption{Examples of local temporal variation of $p^*$ inside the pore, revealing the appearance of oscillations with 2 -- 3 min periods. 
\label{fig:1Dfluct}}
\end{figure}

\begin{figure}
\begin{center}
\includegraphics[scale=1.1]{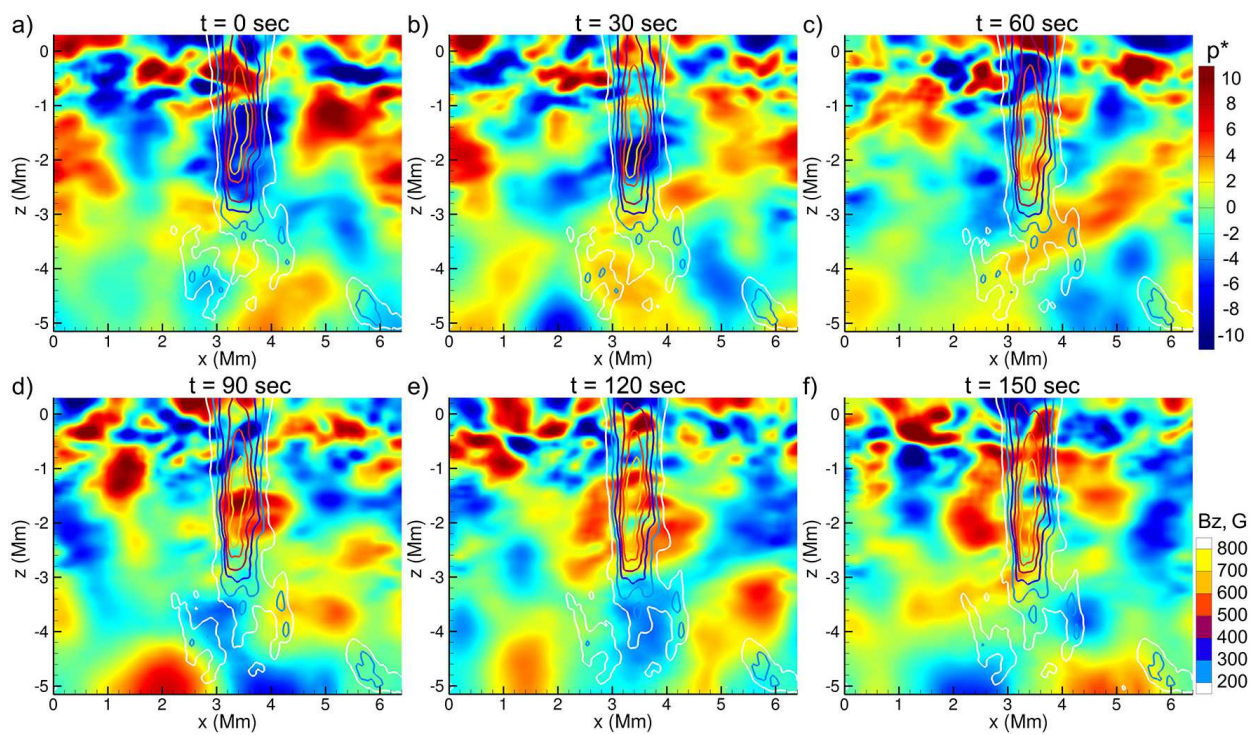}
\end{center}
\caption{Sequence of the time-differences of the gas pressure fluctuations, $p^*$, averaged in the $y$-direction, reveals an acoustic wave excited in the pore. Overplotted isocontours correspond to the vertical magnetic field, averaged in the $y$-direction, in a range from 200~G to 800~G with a 100~G interval. The time-difference between the snapshots is 30~sec.
\label{fig:2Dwave_aver}}
\end{figure}

\begin{figure}
\begin{center}
\includegraphics[scale=0.4]{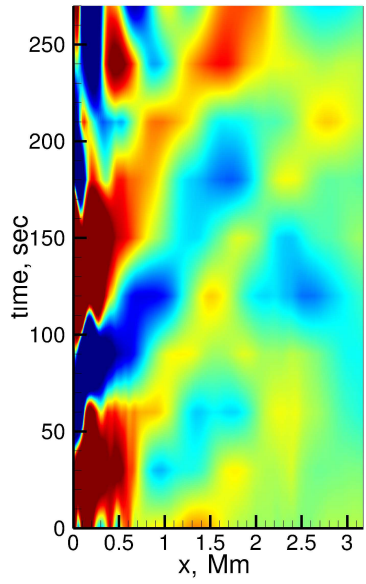} 
\end{center}
\caption{Time-space diagram at depth $\sim 1.2$~Mm of the time-differences of pressure fluctuations shows characteristic ridges that correspond to acoustic waves. 
\label{fig:time-space_diag}}
\end{figure}

The unsteady dynamics of the pore represents strong interactions of several highly twisted magnetic bundles with strengths up to 6 -- 7~kG in the subsurface layers (Fig.~\ref{fig:3D}), which are likely to stabilize the magnetic structure \citep{Parker1979f}. Figure~\ref{fig:2D_wave} shows the evolution of time-differences of the normalized gas pressure, $p^*$ and reveals the appearance of nearly circular patterns centered inside the pore. These regions in the pore "body" experience increases and decreases of $p^*$. To track the local time-evolution of variations of $p^*$ we plot them as a function of time (Fig.~\ref{fig:1Dfluct}), where each plot corresponds to an independent event and shows variations of $p^*$ with a characteristic period of 2 -- 3~min, corresponding to an acoustic wave source. As it seen from Figure ~\ref{fig:2D_wave}, despite the quasiperiodic enhancement of fluctuations inside the pore, it is very difficult to identify wave fronts because of surrounding turbulence. However, averaging the gas pressure perturbations, say in the $y$-direction, across the pore and smoothing can significantly improve the signal-to-noise ratio. Figure~\ref{fig:2Dwave_aver} illustrates wave excitation inside the pore and an expanding perturbation as a ring-like wave pattern. Another way to capture the wave is to plot the time-space diagram. Because the pore evolves in time and also can freely move in the computational domain, we perform tracking of the pore at the photosphere. Then we transform the simulation data to polar coordinates and average over azimuth (Fig.~\ref{fig:time-space_diag}). Note that in most cases it is better to average only over a certain range of azimuth, where turbulent flows are not strong. Because the time-space diagrams can be obtained for each depth, they are also used as an additional criterion for estimating source depth. Figure~\ref{fig:time-space_diag} shows a series of ridges in the time-space diagram that represent propagation of acoustic waves with a speed of 14~km/s about 1.2~Mm below the photosphere.

\begin{figure}
\begin{center}
\includegraphics[scale=1]{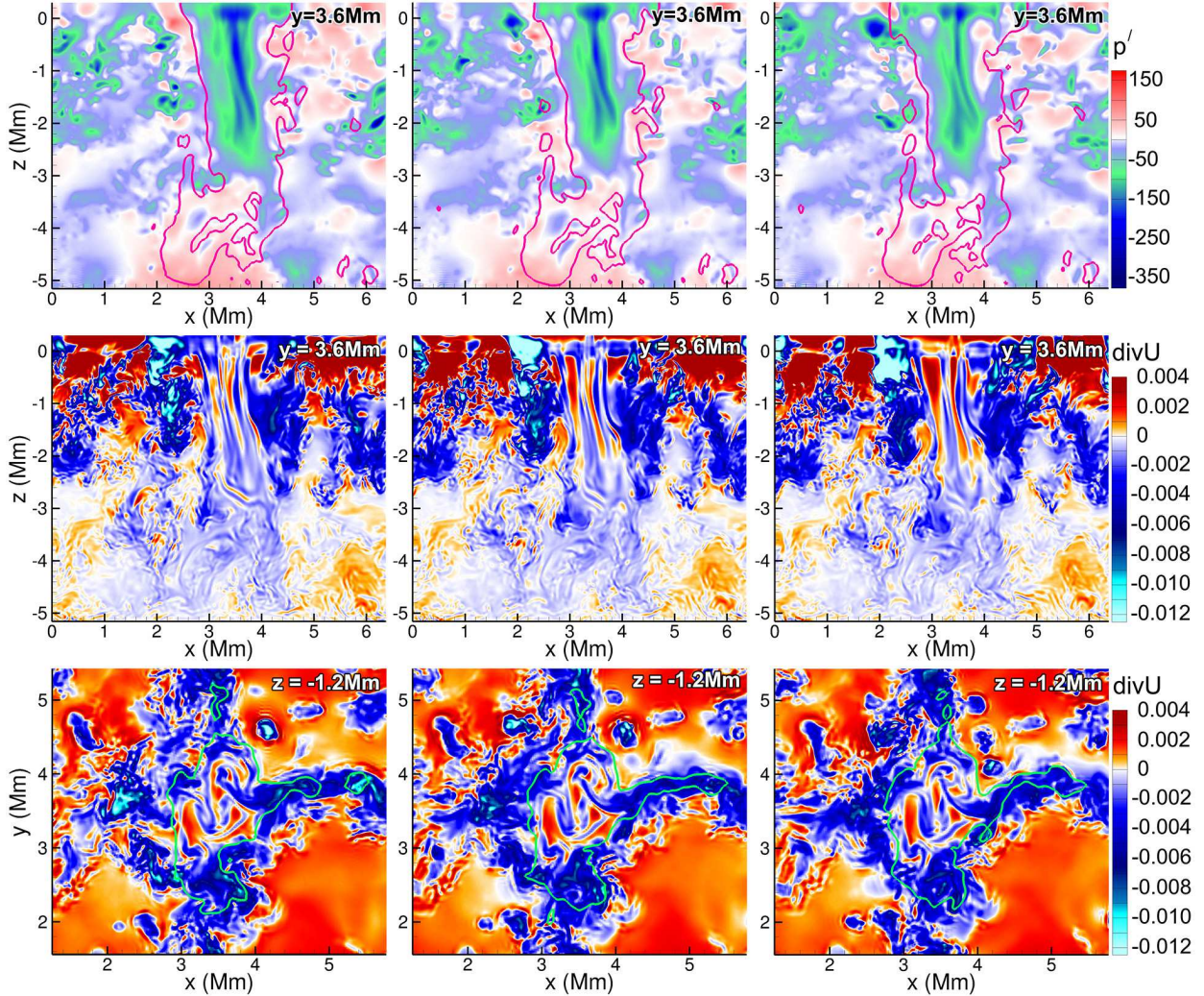}
\end{center}
\caption{Time-evolution (from left to right) of the normalized gas pressure, ${p'}$ (top row), and div(U) (middle and bottom rows) across the pore and a horizontal slice at a depth of $-1.2$~Mm. The images reveal local disturbances and squeezing of the pore by converging flows that lead to excitation of waves inside the pore. The contour lines correspond to a vertical magnetic field of 500~G. Time difference between the snapshots is 1~min.
\label{fig:PprDivU}}
\end{figure}

Our analysis is primarily based on variations of the time-difference of normalized gas pressure fluctuations, $p^*$. The enhancement of this quantity inside the pore (e.g., Fig.~\ref{fig:2D_wave}) reflects local movements of magnetic flux bundles, which form the structure, being pushed by surrounding converging flows. In fact, the converging flows accompany the pore dynamics all the time and maintain its structure. In general, the internal filamentary structure of the pore (Fig.~\ref{fig:3D}) evolves rather slowly, without strong external perturbations from the converging flows of the surrounding plasma. Nevertheless, frequently the converting flows become locally much stronger, squeezing the magnetic bundles and causing strong local variations in the gas pressure. Such local shifts of the bundles are detected in enhancement in the time-difference of gas pressure fluctuation (Fig.~\ref{fig:2D_wave}). The process is illustrated in Figure~\ref{fig:PprDivU} as a time sequence (left to right panels) of the gas pressure perturbations (upper row), $p'$, and the flow divergence, div(U) (middle and bottom rows). The middle column corresponds to the moment of time illustrated in Figure~\ref{fig:2D_wave}a as the time-difference of $p'$. 

The twisting magnetic bundles are dynamically coupled. This makes them more stable so that they evolve more slowly than surrounding flows. The surrounding converging flows (Fig.~\ref{fig:PprDivU}, middle and bottom rows) make such interactions even stronger. Strong local enhancements of converging streams (light blue colors in divU, Fig.~\ref{fig:PprDivU}) in one or more areas initiate local compression of the magnetic structure and its expansion. Because such local impacts quickly come to involve the whole pore, the horizontal location of the compression cannot be determined precisely and certainly cannot be considered as a point source. Nevertheless, the depth of the acoustic source can be determined by tracking the wave fronts back in time to their origin, analyzing the time-differences of gas pressure fluctuations, and using time-space diagrams. In the next section we will use the identified acoustic sources to investigate statistical correlations of various properties.

\section{Statistical properties of acoustic sources}\label{sec:pore-stat}

We separately consider acoustic events beneath the solar surface in areas of strong magnetic field (inside the pore) and in weak fields. To perform a statistical study we consider only the last 5 hours of the data set, to avoid the pore formation stage. From the statistical analysis we exclude acoustic events in the following cases: 1) excited wave interferes with a wave reflected from the bottom boundary; 2) acoustic source is close to the bottom boundary so that is not possible to properly track the wave front; and 3) uncertainty of the source location is greater than 100~km (e.g., due to high concentration of several sources in areas with interacting downdrafts). In this study more than 600 events have been identified, and, in most cases, the localization error does not exceed 50 -- 75 km. Because of this uncertainty, we compare mean physical properties in the vicinity of the source location by averaging over a 100 km$^3$ volume (Fig.~\ref{fig:stat}). Distribution of the identified sources as a function of depth and magnetic field strength shows a clear separation between events associated with the pore dynamics in strong field regions (red dots) and events located outside the pore in areas of weak magnetic field (blue dots). There are few cases of acoustic events associated with the pore dynamics but located in areas with weak magnetic field (Fig.~\ref{fig:stat}a). These exceptions reflect the natural dynamics of the pore filamentary structure, because in layers below 2~Mm the "roots" of the magnetic structure can split into two or more substructures with a weak magnetic field between the bundles.

The relative distribution of the sources in terms of the time-difference of normalized gas pressure perturbations generally reveals stronger fluctuations in the pore (Fig.~\ref{fig:stat}b). Pressure fluctuations in weak field areas are stronger closer to the solar photosphere, and their strength decreases with depth. The sources in the strong magnetic field areas lie in a relatively narrow range of gas pressure fluctuations, while in the weak field areas the distribution of pressure fluctuations is much broader (Fig.~\ref{fig:stat}c). Vertical velocity fluctuations in the vicinity of the acoustic sources located in the pore are about 1~km/s and weakly decrease with the depth. Outside the pore areas, the vertical velocity of the sources has a broad distribution and rapidly decreases with depth (Fig.~\ref{fig:stat}d).

The histogram of the acoustic source distribution with depth shows that most of the sources outside the magnetic structure are located around 0.5~Mm beneath the surface and that the number of excitation events quickly decreases with depth (blue bars, Fig.~\ref{fig:hist}). Acoustic waves inside the magnetic structure are excited in a broad range of depths, from 1 to 2 Mm (red bars). The number of the acoustic events deeper than 2.5~Mm is likely underestimated due to selection effects.

\section{Discussion and conclusion}\label{sec:conclusion}

Recent local helioseismology inferences show signatures of acoustic waves \citep{Zhao2015} excited in sunspot areas at a depth of about 5~Mm, much deeper than was found in previous studies of acoustic wave excitation in quiet-Sun regions \citep[e.g.][]{Goode1992,Kumar1994,Kumar1999a,Nigam1999a,Kitiashvili2011}. In this paper, we used results of 3D radiative MHD simulations to investigate excitation of acoustic waves in a stable, self-organized pore-like magnetic structure, and compare with acoustic sources in weak-field regions outside this structure. The magnetic structure is spontaneously formed from an initial uniformly distributed 100~G vertical magnetic field \citep{Kitiashvili2010}. The mean magnetic field strength of the pore is about 1.5 -- 1.7~kG at the photosphere and reaches 6 -- 7~kG below the solar surface. The subsurface structure of the pore consists of several highly twisted magnetized bundles. The structure is maintained due to surrounding downdrafts converging around the pore. In the analysis we used the last 5 hours of the simulation run when the structure was fully developed and stable.  We performed a case-by-case study and identified more then 600 individual excitation events both inside and outside the pore.

The performed case studies showed that excitation of the acoustic waves outside the pore are associated with convective downdrafts in the intergranular lanes, as was previously suggested. The depth of the excitation depends on the downdraft speed and presence of magnetic field. The sources then often move downward together with the downdrafts, so that the wave front shape can be deformed along the downdrafts. Excitation of acoustic waves inside the pore is associated with interaction of magnetic bundles, when strong converging plasma streams hit the pore boundary causing compression of the bundles, and enhancements of the gas pressure. Several cases of wave excitation identified inside the pore are associated with weak local magnetic field and were located near the photosphere below 2~Mm. The origin of this phenomenon is in the filamentary structure of the pore. The local weak field regions sometimes appear when subsurface layers of the pore expand, and the distance between the magnetic bundles increases. In deeper layers, the bottom part of the pore can split into several "roots". In these cases, we found sources located near bifurcations in the pore's body and likely associated with the bundle dynamics. 

\begin{figure}
\begin{center}
\includegraphics[scale=0.95]{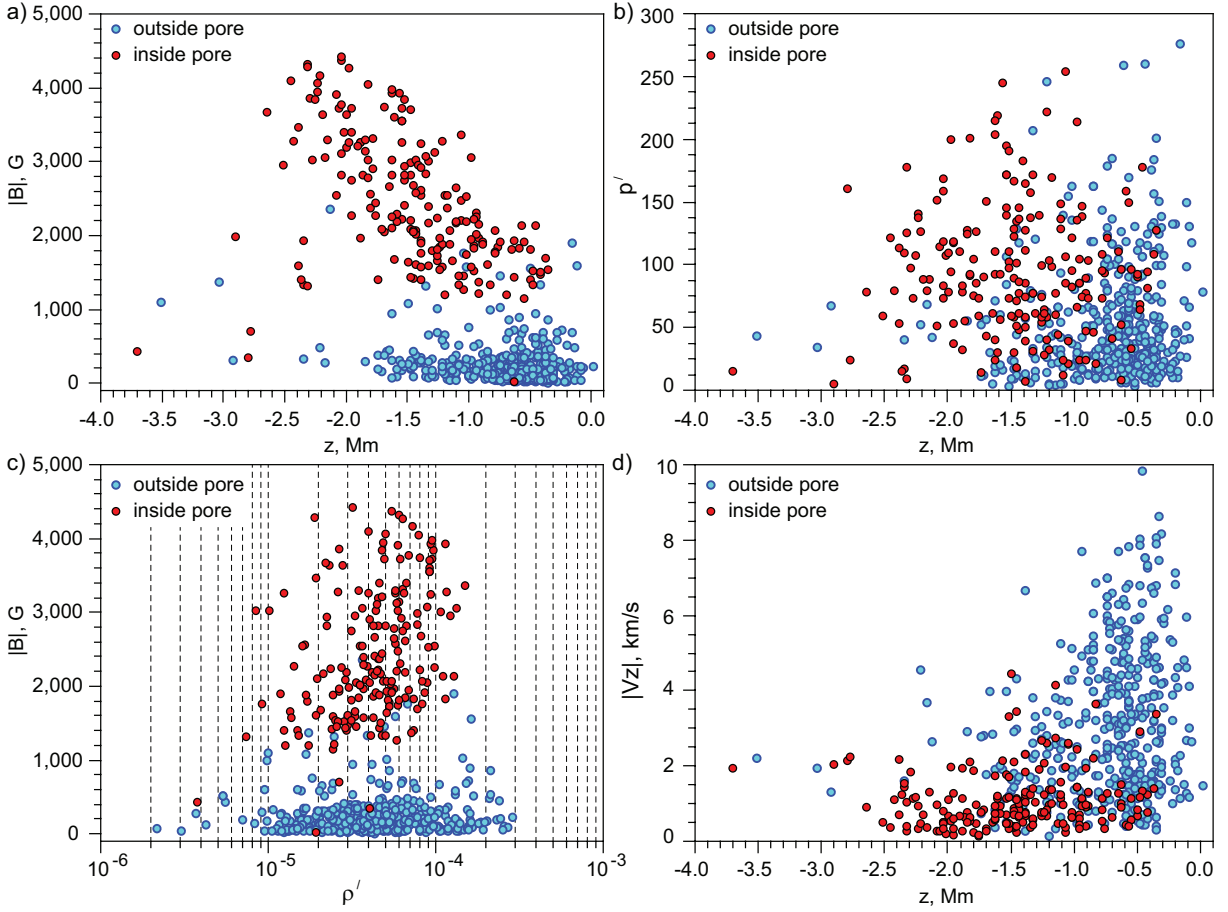}
\end{center}
\caption{Statistical distributions for the acoustic sources located outside (blue circles) and inside (red circles) the pore-like structure. Panels a), b), and d) show scatter plots of the magnetic field strength, normalized gas pressure fluctuations, $p'$, and the vertical velocity magnitude as function of depth in the acoustic source. Panel c) shows the relative distribution of magnetic field and the normalized density fluctuations, $\rho'$.
\label{fig:stat}}
\end{figure}

\begin{figure}
\begin{center}
\includegraphics[scale=1]{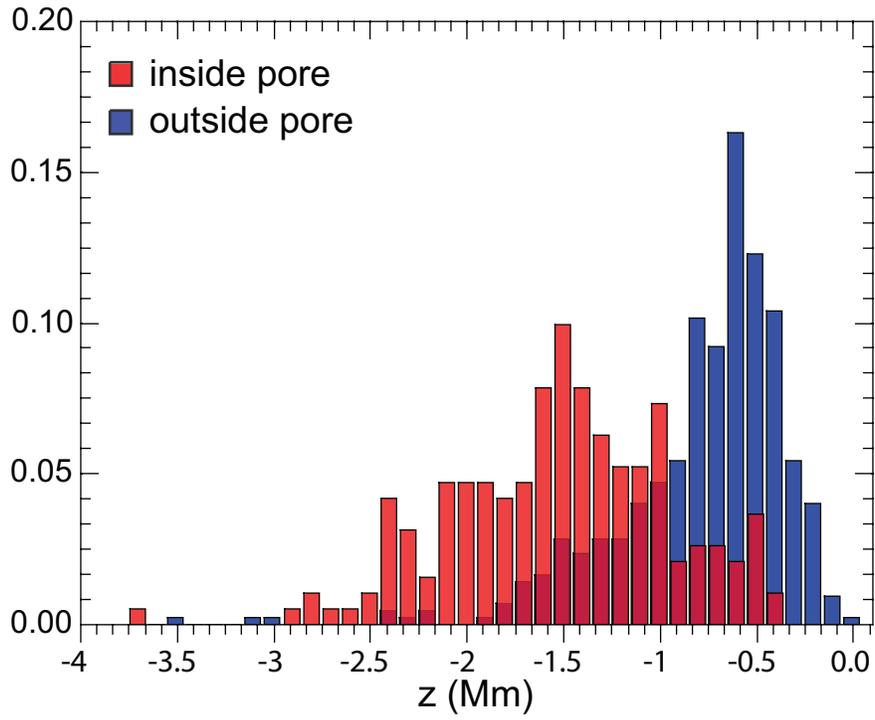}
\end{center}
\caption{Histogram of the acoustic events as a function of the source depth. Red color shows the acoustic source distribution excited inside the magnetic pore. Blue colors correspond to events excited outside the pore.
\label{fig:hist}}
\end{figure}

Statistical distribution of the identified acoustic sources shows a clear separation between the acoustic events outside and inside the pore (Fig.~\ref{fig:hist}). In the area outside the pore, the sources are mostly located in the near-surface layers up to 1~Mm deep, with the distribution peak around 0.5 -- 0.6~Mm. The number of events rapidly decreases with depth. Deeper sources outside the pore are associated with stronger downflows, stronger magnetic fields, and stronger cooling in the intergranular lanes (Fig.~\ref{fig:stat}).

Excitation of acoustic waves in the magnetic pore (or beneath the pore) is driven by the interaction of magnetic bundles. The distribution of acoustic sources with depth is much broader, and most of the identified events are located from 1 to 2.5~Mm below the photosphere (Fig.~\ref{fig:hist}). The maximum number excitation events was around 1.5~Mm below the photosphere. However, the small number of acoustic sources identified below 2.5~Mm is due to limitations of the current numerical simulations. Despite these limitations, our results reveal a new mechanism of acoustic wave excitation in concentrated magnetic structures (observed as pores and sunspots) associated with the interaction of magnetic bundles with deep converging flows that maintain these structures. 
\\
{\bf Acknowledgement.}\\
The research was supported by the NASA Grants NNX12AD05A, NNX14AB68G, NNX16AP05H.


\begin{thebibliography}{}
\expandafter\ifx\csname natexlab\endcsname\relax\def\natexlab#1{#1}\fi

\bibitem[{{Beckers} \& {Schultz}(1972)}]{Beckers1972}
{Beckers}, J.~M., \& {Schultz}, R.~B. 1972, \solphys, 27, 61

\bibitem[{{Bello Gonz{\'a}lez} {et~al.}(2010){Bello Gonz{\'a}lez}, {Flores
  Soriano}, {Kneer}, {Okunev}, \& {Shchukina}}]{BelloGonzalez2010a}
{Bello Gonz{\'a}lez}, N., {Flores Soriano}, M., {Kneer}, F., {Okunev}, O., \&
  {Shchukina}, N. 2010, \aap, 522, A31

\bibitem[{{Bonet} {et~al.}(2008){Bonet}, {M{\'a}rquez}, {S{\'a}nchez Almeida},
  {Cabello}, \& {Domingo}}]{Bonet2008}
{Bonet}, J.~A., {M{\'a}rquez}, I., {S{\'a}nchez Almeida}, J., {Cabello}, I., \&
  {Domingo}, V. 2008, \apjl, 687, L131

\bibitem[{{Bonet} {et~al.}(2010){Bonet}, {M{\'a}rquez}, {S{\'a}nchez Almeida},
  {Palacios}, {Mart{\'{\i}}nez Pillet}, {Solanki}, {del Toro Iniesta},
  {Domingo}, {Berkefeld}, {Schmidt}, {Gandorfer}, {Barthol}, \&
  {Kn{\"o}lker}}]{Bonet2010}
{Bonet}, J.~A., {M{\'a}rquez}, I., {S{\'a}nchez Almeida}, J., {et~al.} 2010,
  \apjl, 723, L139

\bibitem[{Eckart(1960)}]{Eckart1960}
Eckart, C. 1960, Hydrodynamics of Oceans \& Atmospheres, ed. J.~Gillis
  (Pergamon Press)

\bibitem[{{Felipe} \& {Khomenko}(2017)}]{Felipe2017}
{Felipe}, T., \& {Khomenko}, E. 2017, \aap, 599, L2

\bibitem[{{Georgobiani} {et~al.}(2000){Georgobiani}, {Kosovichev}, {Nigam},
  {Nordlund}, \& {Stein}}]{Georgobiani2000}
{Georgobiani}, D., {Kosovichev}, A.~G., {Nigam}, R., {Nordlund}, {\AA}., \&
  {Stein}, R.~F. 2000, \apjl, 530, L139

\bibitem[{{Goldreich} \& {Kumar}(1990)}]{Goldreich1990}
{Goldreich}, P., \& {Kumar}, P. 1990, \apj, 363, 694

\bibitem[{{Goldreich} {et~al.}(1994){Goldreich}, {Murray}, \&
  {Kumar}}]{Goldreich1994}
{Goldreich}, P., {Murray}, N., \& {Kumar}, P. 1994, \apj, 424, 466

\bibitem[{{Goode} {et~al.}(1992){Goode}, {Gough}, \& {Kosovichev}}]{Goode1992}
{Goode}, P.~R., {Gough}, D., \& {Kosovichev}, A.~G. 1992, \apj, 387, 707

\bibitem[{{Hoekzema} {et~al.}(2002){Hoekzema}, {Rimmele}, \&
  {Rutten}}]{Hoekzema2002}
{Hoekzema}, N.~M., {Rimmele}, T.~R., \& {Rutten}, R.~J. 2002, \aap, 390, 681

\bibitem[{{Kitiashvili} {et~al.}(2011){Kitiashvili}, {Kosovichev}, {Mansour},
  \& {Wray}}]{Kitiashvili2011}
{Kitiashvili}, I.~N., {Kosovichev}, A.~G., {Mansour}, N.~N., \& {Wray}, A.~A.
  2011, \apjl, 727, L50

\bibitem[{{Kitiashvili} {et~al.}(2010){Kitiashvili}, {Kosovichev}, {Wray}, \&
  {Mansour}}]{Kitiashvili2010}
{Kitiashvili}, I.~N., {Kosovichev}, A.~G., {Wray}, A.~A., \& {Mansour}, N.~N.
  2010, \apj, 719, 307

\bibitem[{{Komm} {et~al.}(2000){Komm}, {Howe}, \& {Hill}}]{Komm2000}
{Komm}, R.~W., {Howe}, R., \& {Hill}, F. 2000, \apj, 543, 472

\bibitem[{{Kumar}(1994)}]{Kumar1994}
{Kumar}, P. 1994, \apj, 428, 827

\bibitem[{{Kumar} \& {Basu}(1999)}]{Kumar1999a}
{Kumar}, P., \& {Basu}, S. 1999, \apj, 519, 396

\bibitem[{{Kumar} \& {Lu}(1991)}]{Kumar1991}
{Kumar}, P., \& {Lu}, E. 1991, \apjl, 375, L35

\bibitem[{{Musielak} {et~al.}(1994){Musielak}, {Rosner}, {Stein}, \&
  {Ulmschneider}}]{Musielak1994}
{Musielak}, Z.~E., {Rosner}, R., {Stein}, R.~F., \& {Ulmschneider}, P. 1994,
  \apj, 423, 474

\bibitem[{{Nigam} \& {Kosovichev}(1999)}]{Nigam1999a}
{Nigam}, R., \& {Kosovichev}, A.~G. 1999, \apjl, 514, L53

\bibitem[{{Parker}(1979)}]{Parker1979f}
{Parker}, E.~N. 1979, \apj, 230, 905

\bibitem[{{Rast}(1995)}]{Rast1995}
{Rast}, M.~P. 1995, \apj, 443, 863

\bibitem[{{Rast}(1999)}]{Rast1999}
---. 1999, \apj, 524, 462

\bibitem[{{Renaud} {et~al.}(1999){Renaud}, {Grec}, {Boumier}, {Gabriel},
  {Robillot}, {Cort{\'e}s}, {Turck-Chi{\`e}ze}, \& {Ulrich}}]{Renaud1999}
{Renaud}, C., {Grec}, G., {Boumier}, P., {et~al.} 1999, \aap, 345, 1019

\bibitem[{{Schmidt} {et~al.}(1999){Schmidt}, {Stix}, \&
  {W{\"o}hl}}]{Schmidt1999}
{Schmidt}, W., {Stix}, M., \& {W{\"o}hl}, H. 1999, \aap, 346, 633

\bibitem[{{Skartlien} {et~al.}(2000){Skartlien}, {Stein}, \&
  {Nordlund}}]{Skartlien2000}
{Skartlien}, R., {Stein}, R.~F., \& {Nordlund}, {\AA}. 2000, \apj, 541, 468

\bibitem[{{Stein}(1967)}]{Stein1967}
{Stein}, R.~F. 1967, \solphys, 2, 385

\bibitem[{{Strous} {et~al.}(2000){Strous}, {Goode}, \& {Rimmele}}]{Strous2000}
{Strous}, L.~H., {Goode}, P.~R., \& {Rimmele}, T.~R. 2000, \apj, 535, 1000

\bibitem[{{Ulrich}(1970)}]{Ulrich1970}
{Ulrich}, R.~K. 1970, \apj, 162, 993

\bibitem[{{Wray} {et~al.}(2015){Wray}, {Bensassi}, {Kitiashvili}, {Mansour}, \&
  {Kosovichev}}]{Wray2015}
{Wray}, A.~A., {Bensassi}, K., {Kitiashvili}, I.~N., {Mansour}, N.~N., \&
  {Kosovichev}, A.~G. 2015, ArXiv e-prints, arXiv:1507.07999

\bibitem[{{Wray} {et~al.}(2018){Wray}, {Bensassiy}, {Kitiashvili}, {Mansour},
  \& {Kosovichev}}]{Wray2018}
{Wray}, A.~A., {Bensassiy}, K., {Kitiashvili}, I.~N., {Mansour}, N.~N., \&
  {Kosovichev}, A.~G. 2018, Realistic simulations of Stellar Radiative MHD. In
  Book: Variability of the Sun and Sun-like Stars: from Asteroseismology to
  Space Weather, ed. E.~B. J.-P.~Rozelot (EDP Sciences), 39--62

\bibitem[{{Zhao} {et~al.}(2015){Zhao}, {Chen}, {Hartlep}, \&
  {Kosovichev}}]{Zhao2015}
{Zhao}, J., {Chen}, R., {Hartlep}, T., \& {Kosovichev}, A.~G. 2015, \apjl, 809,
  L15

\bibitem[{{Zhao} {et~al.}(2010){Zhao}, {Kosovichev}, \& {Sekii}}]{Zhao2010}
{Zhao}, J., {Kosovichev}, A.~G., \& {Sekii}, T. 2010, \apj, 708, 304
\end{thebibliography}

\newpage

\end{document}